\documentstyle[preprint,aps]{revtex}
\tightenlines 
\begin{document}
\title{Laser Phase Modulation Approaches towards Ensemble Quantum Computing}
\author{Debabrata Goswami} 
\address{
Tata Institute of Fundamental Research, Homi Bhabha Road,
Mumbai 400 005, India.
}
\date{\today}
\maketitle
\begin{abstract}
Selective control of decoherence is demonstrated for a multilevel system by 
generalizing the instantaneous phase of any chirped pulse as individual 
terms of a Taylor series expansion. In the case of a simple two-level 
system, all odd terms in the series lead to population inversion while the 
even terms lead to self-induced transparency. These results also hold for 
multiphoton transitions that do not have any lower-order photon resonance or 
any intermediate virtual state dynamics within the laser pulse-width. Such 
results form the basis of a robustly implementable CNOT gate.

\end{abstract}


Minimizing decoherence is an important challenge towards the realization of 
quantum computing (QC). While Nuclear Magnetic Resonance (NMR) spectroscopy 
of molecules in liquid phase is an important approach to QC \cite{ct1} and has long 
decoherence times, it suffers in terms of scalability to larger number of 
qubits \cite{ct2}. Alternative approaches for QC are, therefore, being sought \cite{ct3}.   
Optical spectroscopy of molecules and nanoparticles could potentially become 
a highly scalable approach, but for the intramolecular relaxation processes, 
such as Intramolecular Vibrational Relaxation (IVR), which is the most 
important contributor to decoherence even in isolated molecules. Controlling 
IVR is also important for ``coherent-control'' due to its connection to 
``bond-selective'' chemistry  \cite{ct4}.  The prospect of QC can lead to a further 
resurgence of activity in this area. Restriction of IVR by optical schemes \cite{ct5} 
can be an attractive route towards selective excitation in large 
molecular systems. Albeit attractive, most of the photon-mediated approaches 
towards restricting IVR (also called ``photon-locking'') have used 
complicated pulse shapes that are yet to be demonstrated in the laboratory 
due to stringent requirements of intensity and precision. 

In this letter, we use simple chirped pulses, which, by contrast, have been 
produced routinely at very high intensities and at various different 
wavelengths for many applications, including selective excitation of 
molecules in coherent control. Such selective optical control over 
molecules in an ensemble is as unique as addressing individual spins in 
a NMR spectrometer and lead to {\it quantum computing in bulk optical 
systems}. The main aim in coherent control has been controlling an 
observable, while in the case of QC individual logic gates would be 
implemented using the principle of controlling observables. The overall 
idea, therefore, is to use shaped-optical pulses, which, on interaction with 
a quantum system, would retain coherence for longer time so that a larger 
number of logic gates are implementable. 

We show that even simple linearly chirped pulses could restrict IVR in 
systems at least as complicated as investigated earlier \cite{ct5}. We prove that 
higher-order chirps that distort the linear chirps experimentally, in fact, 
result in better performance in restricting IVR. In case of two-level 
systems, such shaped-pulses result either in population inversion or 
self-induced transparency. Similar results hold even in the extreme case of 
a two- or multiphoton transition occurring with a chirped pulse, where the 
lower-order photon processes are non-resonant \cite{ct6,ct7}. This makes these 
results more attractive, since the intensity of the laser fields needed to 
reach adiabatic limits often leads to multiphoton processes.

We apply a linearly polarized laser pulse of the form $E\left( {t} \right) = 
\varepsilon \left( {t} \right)e^{i\left[ {\omega \cdot t + \phi \left( {t} 
\right)} \right]} = \varepsilon \left( {t} \right)e^{i\left[ {\omega + \dot 
{\phi} \left( {t} \right)} \right]t}$ to a simple two-level system with 
$|$1$>$$\rightarrow$$|$2$>$ transition, where $|$1$>$ and $|$2$>$ represent 
the ground and excited eigenlevels, respectively, of the field-free Hamiltonian. 
We have $\varepsilon $({\it t}), $\phi $({\it t}) and $\dot {\phi} \left( {t} 
\right)$ as the instantaneous amplitude, phase and frequency-sweep, 
respectively; and $\omega $ is the laser carrier frequency or the center 
frequency for pulsed lasers. If we expand the instantaneous phase function 
of {\it E}({\it t}) as a Taylor series with constants {\it b}$_{n}$, we have

\begin{equation}
\begin{array}{l}
 \phi \left( {t} \right) = b_{0} + b_{1} t + b_{2} t^{2} + b_{3} t^{3} + 
b_{4} t^{4} + b_{5} t^{5} + .... \\ 
 \dot {\phi} \left( {t} \right) = \;\;\;\quad b_{1} + 2b_{2} t + 3b_{3} 
t^{2} + 4b_{4} t^{3} + 5b_{5} t^{4} + .... \\ 
 \dot {\phi} \left( {t} \right) = \sum\limits_{n=1} n  {b_{n}   
t^{\left( {n - 1} \right)}} \\ 
 \end{array}
\end{equation}

To our knowledge this is the first formulation of a general expression for 
the continuously frequency modulated (``chirped") pulses. 
Establishing this generalization enables us to treat all possible chirped 
pulse cases by exploring the effects of each of the terms in Eqn. (1) initially 
for a simple two-level system and then extend it to the multilevel 
situation for a model five-level system of anthracene molecule, which has 
been previously investigated with complicated shaped-pulses \cite{ct5}. We use a 
density matrix approach by integrating the Liouville equation $\frac{{d\rho 
\left( {t} \right)}}{{dt}} = \frac{{i}}{{\hbar} }\left[ {\rho \left( {t} 
\right),H^{FM}\left( {t} \right)} \right]$ for a Hamiltonian in the rotating 
Frequency Modulated (FM) frame of reference. $\rho $({\it t}) is a 2$ \times 
$2 density matrix whose diagonal elements represent populations in the 
ground and excited states and off-diagonal elements represent coherent 
superposition of states. The Hamiltonian for the simple case of a two-level 
system under the effect of an applied laser field can be written in the 
FM frame for N-photon transition \cite{ct7} as, 
$H^{FM} = \hbar \left( {{\begin{array}{*{20}c} {\Delta + N\dot {\phi} 
\left( {t} \right)}  \hfill & {\Omega}  \hfill \\ {\Omega ^{\ast} }  \hfill & {0} \hfill \\
\end{array}} } \right)$.  The time derivative of the instantaneous phase 
function,~$\dot {\phi} \left( {t} \right)$, appears as an additional resonance 
offset, over and above the time-independent multiphoton detuning 
$\Delta = \omega _{R} - N\omega $, while the direction of the field in 
the orthogonal plane remains fixed. We define the multiphoton Rabi 
Frequencies as complex conjugate pairs: $\Omega $(t)=($\mu _{eff}$.
$\varepsilon $({\it t}))$^{N}$/${\hbar} $ and 
$\Omega ^{*}$(t) =($\mu ^{*}_{eff}$.$\varepsilon $({\it t}))$^{ 
N}$/${\hbar} $. For the $|$1$>$$\rightarrow$$|$2$>$ transition, $\omega _{R} = 
\omega _{2} - \omega _{1} $ is the single-photon resonance frequency. We 
have assumed that the transient dipole moment of the individual intermediate 
virtual states in the multiphoton ladder result in an effective transition dipole 
moment, $\mu _{eff}^{N}$, which is a product of the individual N virtual state 
dipole moments $\mu _{n}$, (i.e., $\mu _{eff}^{N} = \Pi _{n}^{N} \mu _{n} $). 
This approximation is particularly valid when intermediate virtual levels are 
non-resonant and as such their multiphoton interaction dynamics can be 
neglected \cite{ct6,ct7}. 

Let us extend the two-level formalism to the multilevel situation involving 
IVR. In the conventional zeroth order description of intramolecular 
dynamics, the system can be factored into an excited state that is 
radiatively coupled to the ground state, and nonradiatively to other bath 
states that are optically inactive (Fig. 1 inset). These ``dark'' states 
have no radiative transition moment from the ground state as determined by 
optical selection rules. They can belong to very different vibrational modes 
in the same electronic state as the ``bright'' state, or can belong to 
different electronic manifolds. These dark states can be coupled to the 
bright state through anharmonic or vibronic couplings. Energy flows through 
these couplings and the apparent bright state population disappears. 
Equivalently, the oscillator strength is distributed among many eigenstates. 
The general multilevel Hamiltonian in the FM frame for an N-photon 
transition (N$ \ge $1), expressed in the zero-order basis set, is:

\begin{equation}
H^{FM} = {\begin{array}{*{20}c}
 {{\begin{array}{*{20}c}
{\hspace{0.5em}\left| {0} \right\rangle}  \hfill & {} \hfill & {\;\hspace{0.6em}\left| {1} 
\right\rangle}  \hfill & {} \hfill & {\;\;\left| {2} 
\right\rangle}  \hfill & {} \hfill &  {\;\hspace{0.0em}\left| {3} \right\rangle 
} \hfill & {\quad \hspace{0.0em}\left| {4} \right\rangle}  \hfill 
& {} \hfill & { \hspace{0.0em}\ldots}  \hfill \\
\end{array}} } \hfill \\

{\left( {{\begin{array}{*{20}c}
 {0} \hfill & {\Omega \left( {t} \right)} \hfill & {0} \hfill & {0} \hfill & 
{0} \hfill & { \ldots}  \hfill \\
 {\Omega ^{\ast} \left( {t} \right)} \hfill & {\delta _{1}
\left( {t} \right)} \hfill & {V_{12}}  \hfill & {V_{13}}  \hfill & { V_{14}} \hfill & { \ldots 
} \hfill \\
 {0} \hfill & {V_{12}}  \hfill & {\delta _{2}  \left( {t} 
\right)} \hfill & {V_{23}}  \hfill & { V_{24}} \hfill & { \ldots}  \hfill \\
 {0} \hfill & {V_{13}}  \hfill & {V_{23}}  \hfill & {\delta _{3} \left( {t} \right)} \hfill & { V_{34}} \hfill & { \ldots}  \hfill \\
 {0} \hfill & {V_{14}}  \hfill & {V_{24}}  \hfill & { V_{34}} \hfill & {\delta _{4} \left( {t} \right)} \hfill & { \ldots}  \hfill \\
 { \vdots}  \hfill & { \vdots}  \hfill & { \vdots}  \hfill & { \vdots}  
\hfill & { \vdots}  \hfill & {} \hfill \\
\end{array}} } \right)} \hfill \\
\end{array}} 
\end{equation}

\noindent
where, $\Omega$(t) (and its complex conjugate pair, $\Omega ^{*}$(t)) is the 
transition matrix element expressed in Rabi frequency units, between 
the ground state $|$0$>$ and the excited state $|$1$>$. The background 
levels $|$2$>$, $|$3$>$,... are coupled to $|$1$>$ through the 
matrix elements V$_{12}$, V$_{23}$, etc. Both the Rabi frequency $\Omega 
$(t) and the detuning frequency [$\delta_{1,2,...}$=$\Delta _{1,2,...}$ + {\it N}$\dot {\phi 
}\left( {t} \right)$] are time dependent (the time dependence is completely 
controlled by the experimenter). In general, the applied field would couple 
some of the dark states together, or would couple $|$1$>$ to dark states, and 
thus, the V$_{ij}$ terms would have both an intramolecular, time independent 
component and a field-dependent component. As an alternative to Eqn. (2), the 
excited states' submatrix containing the bright state $|$1$>$ and the bath 
states $|$2$>$, $|$3$>$,... can be diagonalized to give the eigenstate representation 
containing a set of $\Delta $'$_{i}$ as diagonal elements and corresponding 
$\Omega $'$_{i}$ as off-diagonal elements. Such a representation corresponds 
closely to what is observed in conventional absorption spectroscopy. As long 
as the intensity of the field is very low ($|$$\Omega $'$_{i}$$|$$\ll$$\Delta 
$'$_{i}$) the oscillator strength from the ground state (and hence the 
intensity of the transition, which is proportional to $|\Omega $'$_{i}$ 
$|$$^{2}$) is distributed over the eigenstates, and the spectrum mirrors the 
distribution of the dipole moment. On the other hand, a pulsed excitation 
creates a coherent superposition of the eigenstates within the pulse 
bandwidth. Physically, in fact, the presence of the dark states has been key to 
the loss of selectivity of excitation to a specified bright state. 

From experimental results on the fluorescence quantum beats in jet-cooled 
anthracene \cite{ct8}, the respective values (in GHz) of $\Delta _{1,2,...,4}$ 
are 3.23, 1.7, 7.57 and 3.7; and V$_{12}$= -0.28, V$_{13}$= -4.24, V$_{14}$= -1.86, 
V$_{23}$= 0.29, V$_{24}$= 1.82, V$_{34}$= 0.94. When these values are 
incorporated in Eqn. (2), we obtain the full zero-order Hamiltonian matrix that 
can simulate the experimental quantum beats (Fig. 1) upon excitation with a 
transform-limited Gaussian pulse ($\dot {\phi} \left( {t} \right) = \;0$). Since 
$|$0$>$ and $|$1$>$ do not form a closed two-level system, considerable 
dephasing occurs during the second half of the Gaussian pulse. Thus, in a 
coupled multilevel system, simple pulses cannot be used to generate sequences 
of $\pi $/2 and $\pi $ pulses, as in NMR. The dark states start contributing to 
the dressed state, well before the pulse reaches its peak and results in 
redistributing the population from the from the bright state ($|$1$>$)
into the dark states (Fig. 1). 

A linear sweep in frequency of the laser pulse (i.e., $\dot {\phi} \left( 
{t} \right) = 2b_{2} t$) can be generated by sweeping from far above 
resonance to far below resonance (blue to red sweeps), or its opposite. For 
a sufficiently slow frequency sweep, the irradiated system evolves with the 
applied sweep and the transitions are ``adiabatic''. If this adiabatic 
process is faster than the characteristic relaxation time of the system, 
such a laser pulse leads to a smooth population inversion, i.e., an 
adiabatic rapid passage (ARP) \cite{ct9,ct10}. If the frequency sweeps from below 
resonance to exact resonance with increasing power, and then remains 
constant, adiabatic {\it half} passage occurs and photon locking is achieved 
with no sudden phase shift. However, even under adiabatic {\it full} passage 
conditions, Fig. 2 shows that there is enough slowing down of the {\bf {\it 
E}} field to result in photon locking over the FWHM of the pulse. These 
results hold even under certain multiphoton conditions where only an 
N$^{th}$ (N$ \ge $2) photon transition is possible \cite{ct7}. Theoretically, 
scaling the number of dark states is possible as long as there is finite 
number of states and there are no physical limitations on Stark shifting.

The Quadratic Chirp, i.e., $\dot {\phi} \left( {t} \right) = 3b_{3} t^{2}$, 
is the most efficient in decoupling the bright and dark states as long as 
the Stark shifting of these states prevail at the peak of the pulse. As the 
pulse is turned off, the system smoothly returns to its original unperturbed 
condition (Fig. 3). This would be a very practical approach of controlling 
the coupling of the states with realistic pulse shapes. The cubic term, 
i.e., $\dot {\phi} \left( {t} \right) = \;4b_{4} t^{3}$ behaves more like 
the linear term (Fig. 4). It also decouples the bright and dark states as 
long as the Stark shifting of these states prevail at the peak of the pulse. 
However, the oscillatory nature of the ``photon-locking'' shows that the 
higher-order terms in the Taylor series involve more rapid changes and fails 
to achieve perfect adiabatic conditions. As the pulse is turned off, it 
attempts to invert the bright state population, which quickly dephases, 
analogous to the linear chirp case. Thus, in an isolated two-level system 
that does not suffer from the population dephasing, the linear, cubic, and 
all the higher odd-order terms of the Taylor series (Eqn. (1)) yield 
inversion of population, while the even-order terms produce self-induced 
transparency.  

For a multilevel system, the induced optical AC Stark-shift by the frequency 
swept pulse moves the off-resonant coupled levels far from the resonant 
state leading to an effective decoupling. Under the perfectly adiabatic 
condition, pulses with the even terms in the Taylor series return the system 
to its unperturbed condition at the end. In fact, all higher-order odd terms 
behave in one identical fashion and the even terms behave in another 
identical fashion. It is only during the pulse, that the Stark-shifting of 
the dark states are decoupled and IVR restriction is possible in the 
multi-level situation. In the present calculations, we have used equal 
values to {\it b}$_{n}$ in Eqn. (1), to bring out the effects of the 
higher-order terms in the series. In practice, since Eqn. (1) represents a 
convergent series, only lower-order terms are more important, and since all 
higher-order terms produce the same qualitative results as the lower-order 
terms, one needs to consider only up to the quadratic term. 

The results are generic and illustrate that the intramolecular dephasing can 
be kept to a minimum for the duration of the ``locking'' period under 
adiabatic conditions. Since the effect occurs under an adiabatic condition 
in all these frequency swept pulses, it is insensitive to the inhomogeneity 
in Rabi frequency. The simulations have been performed with laser pulses 
with Gaussian as well as hyperbolic-secant intensity profiles over a range 
of intensities. They show identical results of ``locking'' the population in 
the chosen excited state of a multilevel system, conforming to the adiabatic 
arguments that there is hardly any effect of the actual envelope profile. 
Promoting novel chemical reactions during photon locking, or completing 
several quantum-computing operations can, thus, be accomplished within the 
pulse before dephasing randomizes the initially prepared state. 

In terms of quantum computing, when a sequence of such experimentally 
achievable chirped-pulses act on a bulk system, and perform a series of 
quantum logic gates (e.g., AND, NOT etc.), it would essentially constitute 
``ensemble quantum computing'' \cite{ct11}. Given the formalism of generalized 
chirped pulses discussed in this letter, we demonstrate the construction of 
a CNOT gate as an example of experimentally implementable QC gate involving the
entanglement of coherent photon with quantum system. The truth table for this CNOT gate 
is shown in Table 1 for a quantum mechanical ensemble B that can either be 
in the ground (state 0) or excited (state 1) interacting with the control 
pulse A, which provides robust chirped pulse inversion (condition 1) and the 
self-induced transparency or dark pulse (condition 0). Under the effect of 
inverting pulse the entanglement results in population inversion, while the 
effect of the dark pulse is to preserve the original state of the system. Shaped 
pulses are important for implementing such an ensemble gate since such coupled 
systems preclude the use of simple pulse area approaches of $\pi $/2 and $\pi $ 
sequences for controlling coherence or inducing inversion. 

This straightforward laser phase modulation approach towards a robustly 
implementable ensemble CNOT gate will hopefully form the basis for future 
developments towards ``ensemble quantum computing'' with optical schemes.

\begin{figure}

\caption{ A transform-limited Gaussian pulse interacts with a model 
Anthracene molecule (inset) in a single photon mode or in a multiphoton 
condition. (Inset) Schematic of IVR for Anthracene molecule from Ref. [5] 
based on data extracted from experimental measurements in Ref. [8].}

\caption{ A linearly swept Gaussian pulse (bottom inset, where {\it 
b}$_{2}$=10cm$^{-2}$) can generate ``photon-locking''. The evolution of the 
dressed state character is unchanged while locking occurs (top inset) but as 
the pulse is turned off, the eigen-energy curves cross and the bright state 
population quickly dephases.}

\caption{ The Quadratic Chirped Gaussian pulse (bottom inset, where {\it 
b}$_{3}$=10cm$^{-2}$) is the most efficient in decoupling the bright and 
dark states during the pulse. The eigen-energy curves and the corresponding 
evolution of the dressed state character (top inset) shows that the entire 
process is highly adiabatic.}

\caption{ Effect of a Cubic Chirped Gaussian pulse (bottom inset, where 
{\it b}$_{4}$=10cm$^{-2}$) is similar to the linearly swept pulse (Fig.2), 
although the evidence of population oscillation indicates that this chirp is 
not as adiabatic as the linear chirp. The eigen-energy curves cross towards 
the end of the pulse (top inset) and the bright state population gets 
redistributed.}

\end{figure}
\begin{center}
\begin{table}
\caption{CNOT Gate with Simple Chirped Pulses. 
\label{table1}}
\vspace{1em}
\begin{tabular}{|c|r|r|r|}
Shaped-Pulse& A& B& A$ \oplus $B \\
\tableline
{``Inverting'' Pulse}& 1& 1& 0 \\
\cline{2-4} & 1& 0& 1 \\
\tableline
{``Dark'' Pulse}& 0& 1& 1 \\
\cline{2-4} & 0& 0& 0 \\
\end{tabular}
\end{table}
\end{center}

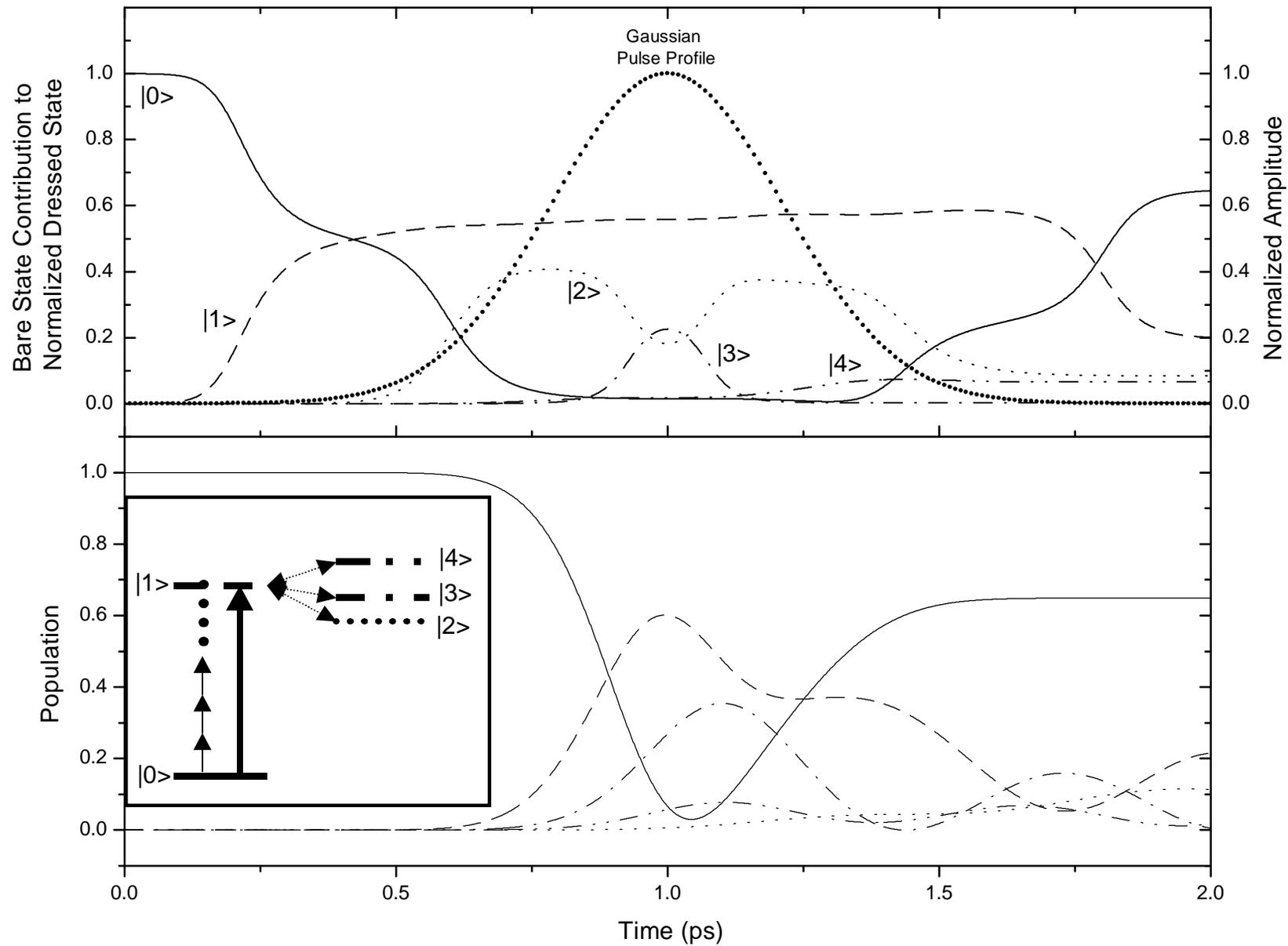
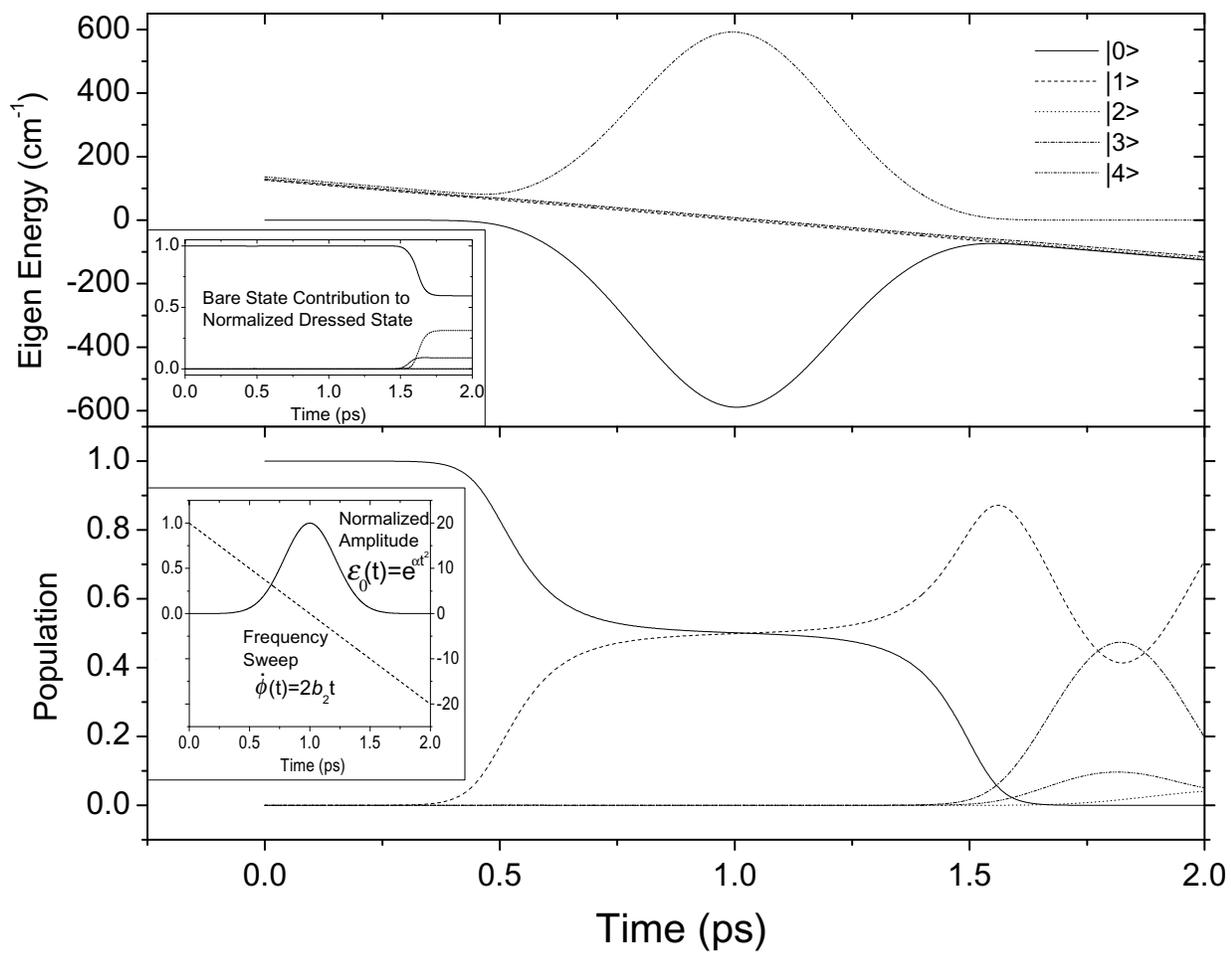
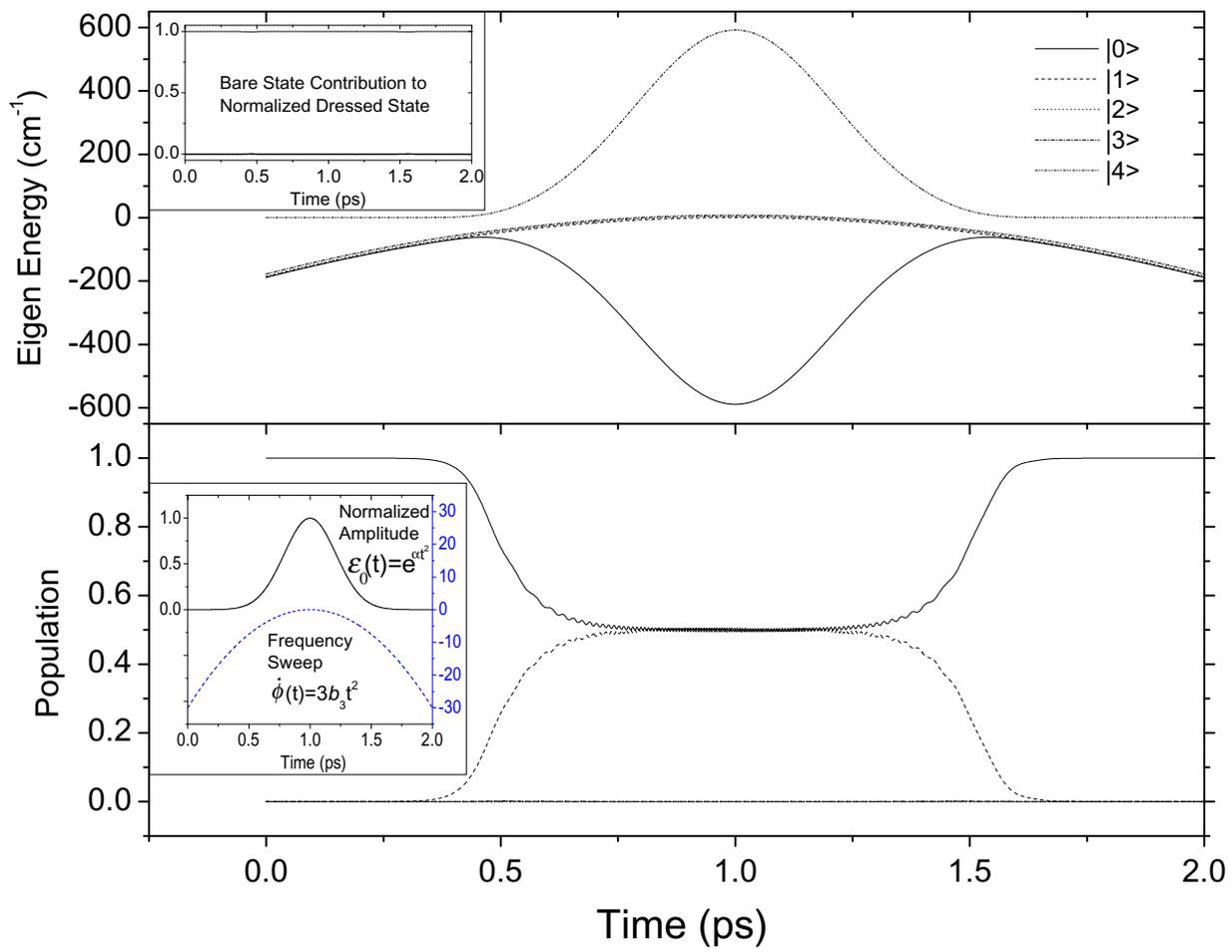
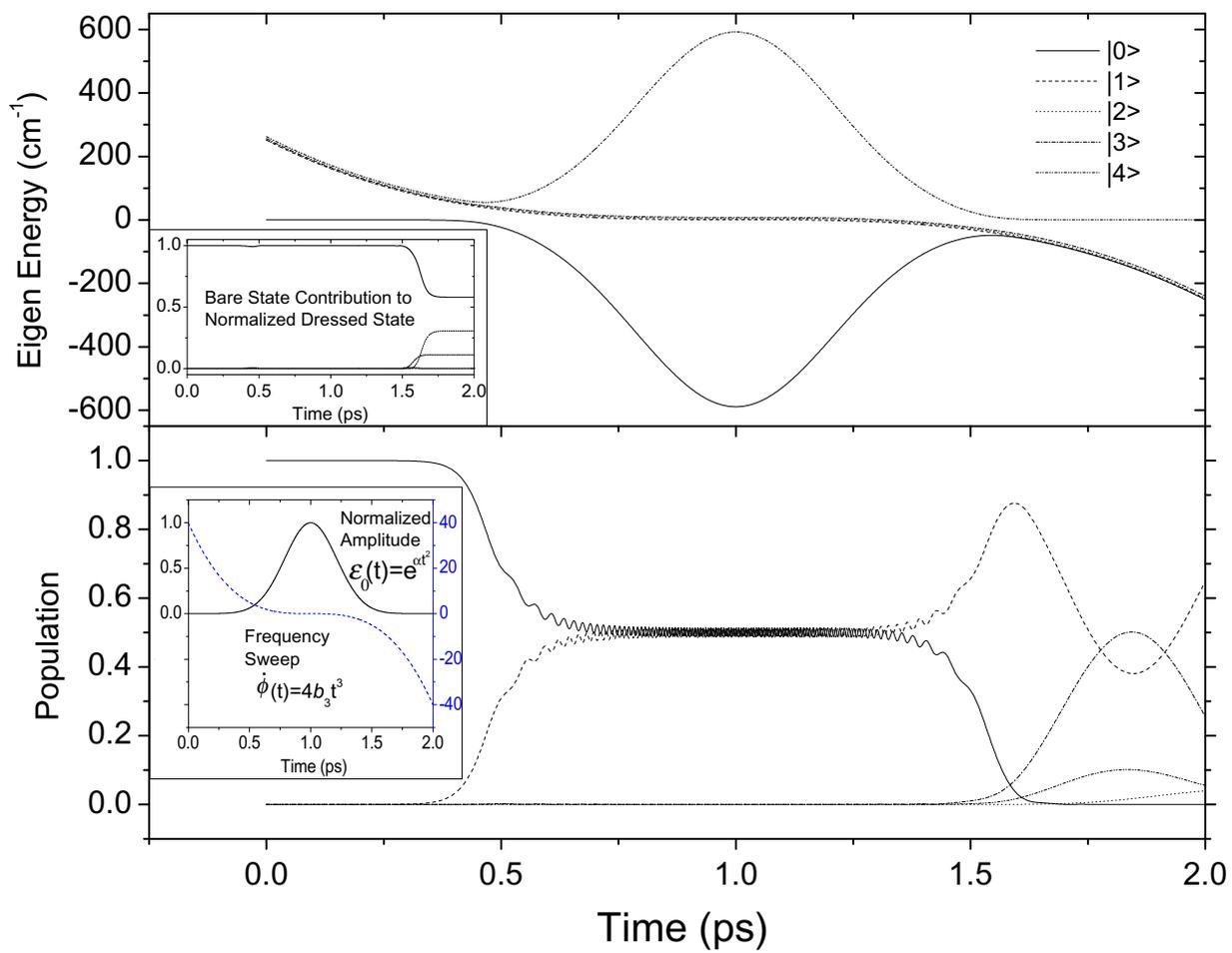
\end{document}